\documentclass[sigconf]{acmart}
\AtBeginDocument{%
  }

\setcopyright{acmlicensed}
\copyrightyear{2025}
\acmYear{2025}
\setcopyright{acmlicensed}\acmConference[RecSys '25]{Proceedings of the Nineteenth ACM Conference on Recommender Systems}{September 22--26, 2025}{Prague, Czech Republic}
\acmBooktitle{Proceedings of the Nineteenth ACM Conference on Recommender Systems (RecSys '25), September 22--26, 2025, Prague, Czech Republic}
\acmDOI{10.1145/3705328.3748080}
\acmISBN{979-8-4007-1364-4/2025/09}

\usepackage[linesnumbered,lined]{algorithm2e}
\usepackage{makecell}
\usepackage{booktabs}
\usepackage{xspace}
\usepackage{multirow}
\usepackage{graphicx}
\usepackage{subfigure}
\usepackage{amsmath}
\usepackage{color}
\usepackage{xcolor}
\usepackage{enumitem}
\usepackage{bbold}

\begin{document}

\title[Exponential-Gaussian Mixture Network for Video Watch Time Prediction]{Multi-Granularity Distribution Modeling for Video Watch Time Prediction via Exponential-Gaussian Mixture Network}

\author{Xu Zhao}
\affiliation{%
  \institution{Xiaohongshu}
  \city{Beijing}
  \country{China}
}
\email{zhaoxu2@xiaohongshu.com}
\orcid{0000-0001-5146-5789}

\author{RuiBo Ma}
\affiliation{%
  \institution{Xiaohongshu}
  \city{Beijing}
  \country{China}
}
\email{maruibo@xiaohongshu.com}
\orcid{0009-0006-4995-0789}

\author{Jiaqi Chen}
\affiliation{%
  \institution{Xiaohongshu}
  \city{Beijing}
  \country{China}
}
\email{chenjiaqi3@xiaohongshu.com}
\orcid{0009-0004-4874-177X}

\author{Weiqi Zhao}
\affiliation{%
  \institution{Xiaohongshu}
  \city{Beijing}
  \country{China}
}
\email{zhanganhao@xiaohongshu.com}
\orcid{0000-0003-2735-408X}

\author{Ping Yang}
\affiliation{%
  \institution{Xiaohongshu}
  \city{Beijing}
  \country{China}
}
\email{yechen1@xiaohongshu.com}
\orcid{0009-0006-2642-3652}

\author{Yao Hu}
\authornote{Yao Hu is the corresponding author.}
\affiliation{%
  \institution{Xiaohongshu}
  \city{Beijing}
  \country{China}
}
\email{xiahou@xiaohongshu.com}
\orcid{0009-0006-1274-7111}

\renewcommand{\shortauthors}{Xu Zhao et al.}

\newcommand{\name}{EGMN\xspace}
\newcommand{\disname}{EGM distribution\xspace}

\begin{abstract}
Accurate watch time prediction is crucial for enhancing user engagement in streaming short-video platforms, although it is challenged by complex distribution characteristics across multi-granularity levels. Through systematic analysis of real-world industrial data, we uncover two critical challenges in watch time prediction from a distribution aspect: \textit{(1)} coarse-grained skewness induced by a significant concentration of quick-skips\footnotemark, \textit{(2)} fine-grained diversity arising from various user-video interaction patterns. Consequently, we assume that the watch time follows the \textbf{E}xponential-\textbf{G}aussian \textbf{M}ixture (EGM) distribution, where the exponential and Gaussian components respectively characterize the skewness and diversity. Accordingly, an \textbf{E}xponential-\textbf{G}aussian \textbf{M}ixture \textbf{N}etwork (\name) is proposed for the parameterization of \disname, which consists of two key modules: a hidden representation encoder and a mixture parameter generator. We conducted extensive offline experiments on public datasets and online A/B tests on the industrial short-video feeding scenario of \href{https://xiaohongshu.com/}{Xiaohongshu App} to validate the superiority of \name compared with existing state-of-the-art methods. Remarkably, comprehensive experimental results have proven that \name exhibits excellent distribution fitting ability across coarse-to-fine-grained levels. We open source related code on Github: \url{https://github.com/BestActionNow/EGMN}.
\end{abstract}



\maketitle

\section{INTRODUCTION}
\label{sec:intro}

\begin{figure*}
    \centering
    \scalebox{1.0}{
    \includegraphics[width=\linewidth]{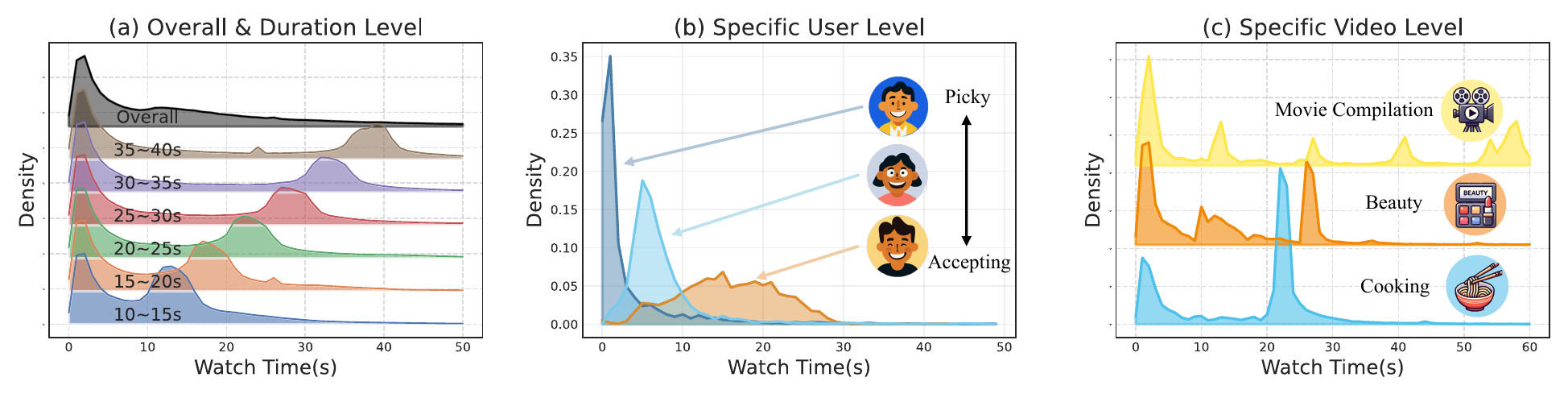}}
        \caption{Investigation of watch time distribution following the coarse-to-fine paradigm in our real industrial scenario.}
    \label{fig:intro}
\end{figure*}

Short-video platforms such as TikTok and KuaiShou have experienced a significant surge in popularity over recent years \cite{savic2021research,wang2019causes,cai2023reinforcing,pan2023understanding}. These services typically employ a continuous swipe-based streaming play pattern without the need to click through multiple candidates explicitly \cite{gao2022kuairand,gong2022real}. Within this paradigm, watch time has emerged as a fundamental indicator of user satisfaction. Therefore, accurate watch time prediction is crucial \cite{wu2018beyond,yi2014beyond,wang2020capturing}, as it empowers the platform to recommend videos that better align with user interests and ultimately enhance user engagement.

The watch time prediction is fundamentally a regression problem since its values are continuous and span a wide range. It is well-known that the distribution of labels significantly influences the difficulty of the regression task, where appropriate distribution assumptions can enhance regression accuracy (as demonstrated by extensive existing studies \cite{ferrari2004beta,ouma2016poisson,wang2019deep,mardalena2020parameter,shono2008application}). However, the distribution of short-video watch time remains insufficiently explored. To bridge this knowledge gap, we thoroughly investigate the distribution of short-video watch time using our real-world industrial recommendation data. As shown in Figure~\ref{fig:intro}, we analyze and compare watch time distributions across coarse-to-fine-grained levels: \footnotetext{Quick-skip indicates that users quickly lose interest and swipe away the video.}
\begin{itemize}[leftmargin=*,topsep=3pt,itemsep=3pt]

\item Overall Level: We examine the overall distribution of watch time, which reveals pronounced skewness around zero, as shown in the top line of Figure~\ref{fig:intro}(a).
\item Duration Level: Figure~\ref{fig:intro}(a) also presents the distribution at the duration\footnote{Duration refers to the video length.} level, where a bimodal pattern is observed across different duration groups.
\item Specific User Level: We analyze the watch time distribution of specific users as shown in Figure~\ref{fig:intro}(b), which demonstrates notable diversity on such a fine granularity. Some users are rather picky, quickly swiping past videos until finding one of interest, resulting in a higher skewness of their watch time distribution. Conversely, some users are more accepting, tending to watch each system-recommended video with greater patience, leading to a lower skewness. 
\item Specific Video Level: The watch time distribution of different videos  demonstrates a multimodal phenomenon, and the differences between videos are even more pronounced. Figure~\ref{fig:intro}(c)  provides three representative examples. The cooking video exhibits a bimodal distribution: disinterested users exit early, whereas engaged viewers complete the content. The beauty video’s engagement drives not only partial/full views but also notable replays, yielding a trimodal pattern. In contrast, the movie compilation's segmented structure features common exit points at scene transitions, ultimately manifesting a multimodal distribution.
\end{itemize}
According to the analysis above, we distill two critical challenges for short-video watch time prediction :
\begin{itemize}[leftmargin=*,topsep=2pt]
    \item \textbf{Coarse-grained distribution skewness} arises from a substantial concentration of quick-skips, which requires specialized modeling approaches capable of addressing this localized clustering phenomenon. 
    \item \textbf{Fine-grained distribution diversity} introduces cross-granular incompatibility that amplifies prediction complexity, demanding adaptive architectures to capture heterogeneous characteristics.
\end{itemize}
Existing watch time prediction methods typically circumvent these challenges through two principal approaches. The first approach is label normalization \cite{zhan2022deconfounding,zhao2023uncovering,zhao2024counteracting}, which not only simplifies the label distribution for easier fitting but also provides an unbiased reflection of user interest. However, it may lead to a loss of absolute watch time information, resulting in reduced prediction accuracy. The second approach is task transformation \cite{lin2023tree,sun2024cread}, where the regression task is transformed into a series of classifications. The benefit is that each subclassification task is easier to learn than the overall regression problem, while the process of discretization and subsequent reconstruction inevitably introduces additional errors. 

To mitigate the two challenges mentioned above, and inspired by previous works in other fields \cite{ferrari2004beta,ouma2016poisson,wang2019deep}, we propose to directly regress the absolute values of watch time based on a sound distribution assumption, thereby improving estimation accuracy. 
To reconcile distributional discrepancies across different granularities, we assume that short-video watch time is governed by a latent mixture model, formally defined as the \textbf{E}xponential-\textbf{G}aussian \textbf{M}ixture distribution (\disname), where the exponential component addresses coarse-grained distribution skewness and the Gaussian components adaptively capture fine-grained distribution diversity.

To estimate the parameters of \disname, we propose the corresponding \textbf{E}xponential-\textbf{G}aussian \textbf{M}ixture \textbf{N}etwork (\name) model using a neural network architecture. Initially, we generate a hidden representation shared across all distribution components. Subsequently, the parameters of each distribution component are estimated based on the hidden representation, and a gated network is applied to perform the weighted mixing of multiple distributions. A combination of loss functions is used to optimize our model, each addressing different aspects of the prediction task. Ultimately, \name produces a parameterized instance of \disname in an end-to-end manner and uses its expected value as the final prediction.

Through extensive experiments, we demonstrate the superiority of \name compared to existing state-of-the-art methods. Moreover, we design an ablation study to validate the distributional assumptions and quantify the contribution of each component to our loss function. Comprehensive experimental results have proven that \name exhibits an excellent distribution-fitting ability across coarse-to-fine-grained levels. Remarkably, \name successfully captures the latent interaction effects between user preferences and video characteristics, which are crucial for recommender systems.

In summary, the major contributions of this paper are as follows:
\begin{enumerate}[leftmargin=*,topsep=3pt,itemsep=3pt]
\item Based on our observations of short-video watch time distribution at various granularities, we introduce a well-founded distributional assumption: \disname, which is compatible with the distribution characteristics of each granularity.
\item To parameterize \disname,  we propose \name for watch time prediction, which simultaneously models the concentrated characteristics of quick-skips and captures the latent viewing patterns arising from user-video interactions.
\item Extensive experiments are conducted on both offline datasets and the online short-video feeding scenario of Xiaohongshu App, demonstrating that \name significantly outperforms existing state-of-the-art methods.
\end{enumerate}

\section{RELATED WORK}
\label{sec:related}

Watch time prediction is a critical module in short-video recommender systems, as it serves as a direct and measurable indicator of user engagement. Accurately estimating watch time helps platforms optimize content delivery, improve user satisfaction, and maximize retention. However, traditional approaches to regression-based watch time prediction rely on specific assumptions on label distribution or error distribution. Such as Value Regression (VR) evaluated via Mean Squared Error (MSE) assumes that the watch time conform to the normal distribution, which is violated in real-world scenarios. Watch time data typically follows a highly skewed, long-tailed distribution, with most data being short and only a few lasting significantly longer. This mismatch between modeling assumptions and empirical data can lead to suboptimal predictions, particularly for outliers. Weighted Logistic Regression (WLR)\cite{covington2016deep} fits a weighted logistic regression model, where click samples are weighted by watch-time and unclick samples receive unit weight. Due to the lack of explicit click behaviors in streaming short-video platforms, WLR can not be directly applied to our system. 

Recent research has shifted focus toward watch time debiasing, aiming to extract unbiased user interests from noisy observations. For example, D2Q\cite{zhan2022deconfounding} mitigates duration bias in watch time prediction by reformulating the task as a quantile regression problem, where videos are first partitioned into discrete duration-based bins, and a regression model is trained for each group to estimate its conditional watch time distribution. Although this approach effectively addresses systemic biases across duration, its reliance on hard grouping introduces fundamental limitations. By enforcing rigid duration boundaries, D2Q fails to capture finer-grained behavioral patterns and produces discontinuous predictions across bin edges. 

Another line of work, such as D$^2$CO\cite{zhao2023uncovering}, identifies noisy watching behavior as a confounding factor that distorts true engagement signals. While  D$^2$CO employs a Gaussian Mixture Model (GMM) to separate meaningful watch time from noise, it fails to account for the rich diversity in user-video interactions, such as quick-skips, replays. Meanwhile, CWM\cite{zhao2024counteracting} points out that the duration bias is derived from the truncation of the user’s theoretical maximum watch time by video duration and a cost-based correction function is defined to transfer original watch time into an unbiased regression space. However, CWM's duration truncation neglects replay behaviors, and the transformation of original watch time may cause distribution information loss.

Alternatively, some methods reformulate watch time prediction as a series of classification tasks, preserving ordinal relationships between different watch time. TPM\cite{lin2023tree} leverages a tree-based probability model to capture conditional dependencies between these classifications, improving hierarchical reasoning. Given the highly imbalanced nature of watch time data, CREAD\cite{sun2024cread} conducts a systematic study on how discretization granularity affects both model learning and prediction restoration. It proposes an error-adaptive discretization mechanism to balance bias and variance, yet it lacks deeper analysis of fine-grained patterns. 

Finally, GR\cite{ma2025sequence} presents a Generative Regression framework capable of modeling watch time alongside other continuous variables. GR reformulates watch time prediction as
a sequence generation task and offers a more flexible and expressive approach. Despite these advances, challenges remain in real-time industrial deployment, causal interpretation of watch behaviors and reliance on complex discretization strategy.
\section{PROPOSED METHOD}

\subsection{Problem Definition}

Let $\mathbf{x} \in \mathbb{R}^d$ represent the embedded feature vector for a user-video pair, incorporating user features, video features, and context information, while $t \in \mathbb{R}^{+}$ represent the watch time. Watch time prediction aims to find a function $f(\cdot)\colon \mathbb{R}^d \to \mathbb{R}^{+}$, such that the discrepancy between the predicted outputs $f(\mathbf{x})$ and the ground truth $t$ is minimized under certain metrics.
Previous approaches typically select a specific metric (e.g. MSE) to construct the loss function between $f(\mathbf{x})$ and $t$, and then optimize $f(\cdot)$ through parameter estimation. However, these metrics rely on oversimplified assumptions on the watch time probability distribution $p(t)$ and ignore the inherent heterogeneity across multi-granularity levels, which motivates our probabilistic modeling of $p(t)$ instead of arbitrary function optimization.


\subsection{An Empirical Assumption to $p(t)$}

Inspired by the distribution patterns and challenges discussed in Section~\ref{sec:intro}, we assume that $p(t)$ follows a latent \disname as a mixture of one exponential distribution and $K$ Gaussian distributions, whose density function is formulated as:

%
\begin{equation}
\label{equ:egmd}
p(t) = \omega_0f_{\text{exp}}(t|\lambda) + \sum_{k=1}^{K} \omega_kf_{\text{gauss}}(t|\mu_k, \sigma_k^2)
\end{equation}

where:
\begin{itemize}[leftmargin=*,topsep=4pt]

    \item $f_{\text{exp}}(t|\lambda) = \lambda e^{-\lambda t}$ is the probability density function of an exponential distribution with rate parameter $\lambda$.
    \item $f_{\text{gauss}}(t|\mu, \sigma^2) = \frac{1}{\sqrt{2\pi\sigma^2}}e^{-\frac{(t-\mu)^2}{2\sigma^2}}$ is the probability density function of a Gaussian distribution with mean $\mu$ and variance $\sigma^2$.
    \item $\omega_k$ represents the mixture weight for component $k$, with a sum constraint: $\sum_{k=0}^{K} \omega_k = 1$.
\end{itemize}

Here, we present the assumed distribution selection rationale. The components of \disname are carefully selected based on the observed characteristics of watch time distribution:

\begin{enumerate}[leftmargin=*,topsep=4pt]
    \item \textbf{Exponential Component}: At the coarse-grained level, the distribution is highly skewed, with a significant concentration of quick-skips. The exponential distribution is well-suited for modeling the quick-skipping behavior due to its memoryless property and concentration of probability mass near zero \cite{balakrishnan1996exponential}.
    \item \textbf{Gaussian Components}: At the fine-grained level, the distribution becomes more intricate. Users with diverse viewing habits interact with various types of videos, resulting in a spectrum of latent consumption patterns. Given that Gaussian mixture distributions have been theoretically established as statistically consistent estimators for complex multimodal distributions \cite{mclachlan2019finite,zhuang1996gaussian}, we leverage them to effectively capture diversity in user preferences and video characteristics.
\end{enumerate}

\subsection{Our Approach: \name}

In this section, we propose a model \name for the parameterization of the \disname as illustrated in Figure~\ref{fig:dar}. Specifically, we employ a deep feed-forward network \cite{papamakarios2021normalizing,bishop1994mixture,kairouz2016discrete} that takes the feature vector $\mathbf{x}$ as input and estimates the parameters of \disname.

\subsubsection{\textbf{Hidden Representation Encoder}}

For each user-item pair, we collect features from multiple sources:
\begin{itemize}[leftmargin=*,topsep=2pt]
    \item User features: demographic information, historical behavior patterns, and engagement statistics
    \item Video features: content characteristics, duration, category, and creator information
    \item Contextual features: time of day, weekday patterns and device type
\end{itemize}

Available features are processed through embedding layers to create the feature vector $\mathbf{x}$. Initially, $\mathbf{x}$ is fed into a feature encoder backbone to obtain a hidden representation $\mathbf{h}$, which is shared across multiple components in \disname:
\begin{equation}
\mathbf{h} = g_{\text{backbone}}(\mathbf{x})
\end{equation}
where $g_{\text{backbone}}$ can be instantiated with any feature encoding backbone adaptable to recommender prediction scenarios, e.g. DCN \cite{wang2017deep}, DIN \cite{zhou2018deep}, SENet \cite{hu2018squeeze}, Transformer \cite{vaswani2017attention}, etc.
It should be noted that \name has the potential to be enhanced by specific model design and feature engineering, since \name is proposed as a backbone-agnostic paradigm that inherently supports the integration of user segmentation and contextual signals. 
\begin{figure}
\setlength{\abovecaptionskip}{15pt}
\setlength{\belowcaptionskip}{-10pt}
    \centering
    \scalebox{0.95}{
    \includegraphics[width=\linewidth]{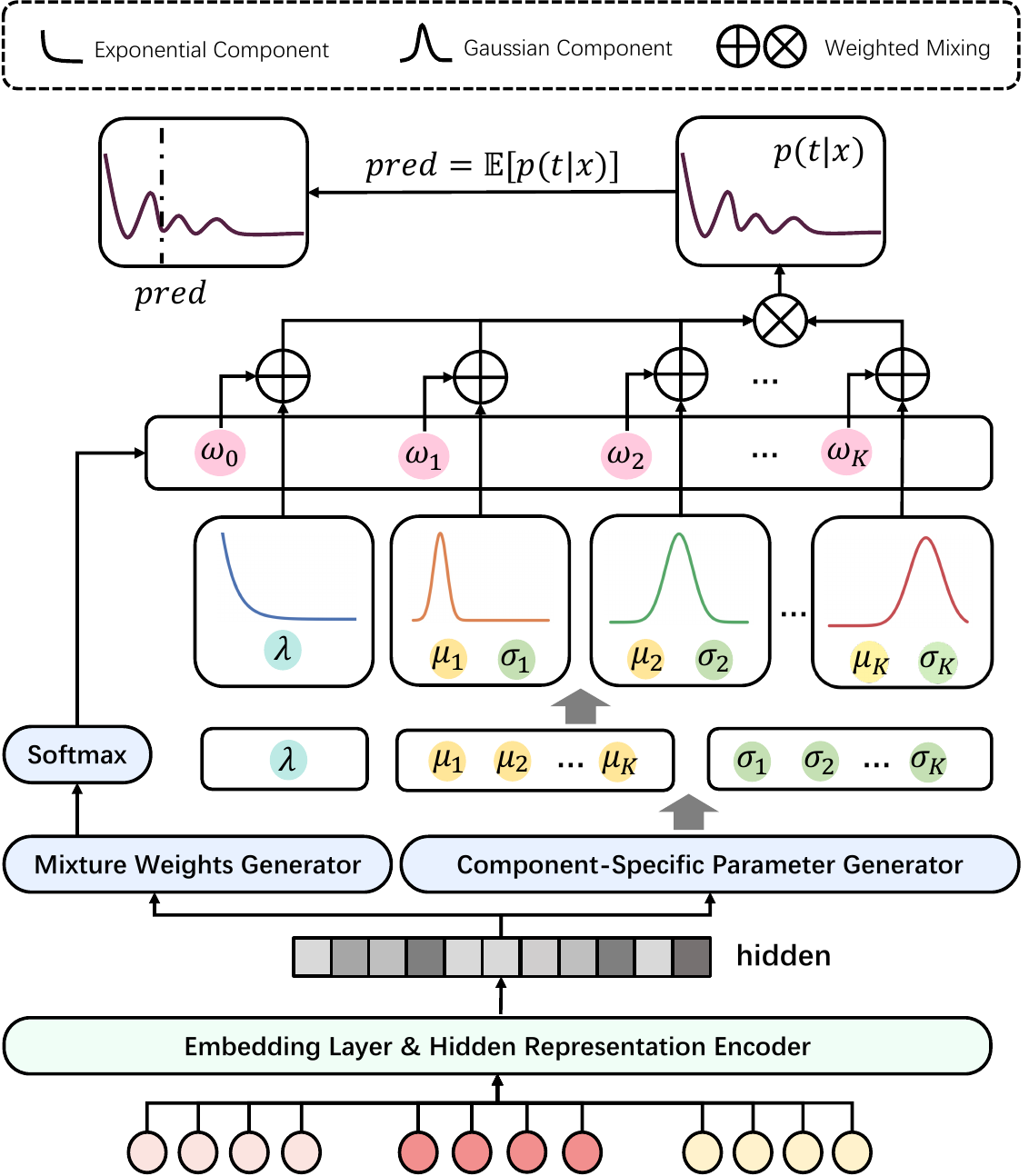}}
    \caption{The proposed \name framework.} 
    \label{fig:dar}
    \vspace{0.2cm}
\end{figure}

\subsubsection{\textbf{Mixture Parameter Generator}} The hidden representation is subsequently fed into separate branches to estimate the parameters of each distribution component:

\begin{enumerate}[leftmargin=*,topsep=4pt]
    \item Exponential Rate: 
    \begin{equation}
    \lambda(\mathbf{x}) = \text{softplus}(\mathbf{W}_{\lambda}\mathbf{h} + \mathbf{b}_{\lambda})
    \end{equation}
    
    \item Gaussian Means: 
    \begin{equation}
    \label{equ-mean}
    \mu_k(\mathbf{x}) = \frac{1}{\lambda(\mathbf{x})} +  \text{softplus}(\mathbf{W}_{\mu_k}\mathbf{h} + \mathbf{b}_{\mu_k})
    \end{equation}
    for $k \in \{1, \ldots, K\}$
    
    \item Gaussian Variances: 
    \begin{equation}
    \sigma_k^2(\mathbf{x}) = \text{softplus}(\mathbf{W}_{\sigma_k}\mathbf{h} + \mathbf{b}_{\sigma_k})
    \end{equation}
    for $k \in \{1, \ldots, K\}$
    
    \item Mixture Weights: 
    \begin{equation} 
    [\omega_0(\mathbf{x}), \omega_1(\mathbf{x}), \ldots, \omega_K(\mathbf{x})] = \text{softmax}(\mathbf{W}_{\omega}\mathbf{h} + \mathbf{b}_{\omega})
    \end{equation}
\end{enumerate}

In formulas above, $\mathbf{W}_{*}$ and $\mathbf{b}_{*}$ are learnable parameters in linear transformation.
We use softplus activation ($\text{softplus}(z) = \log(1 + e^z)$) for the rate, mean and variance parameters to ensure positivity, and softmax activation for the mixture weights to guarantee their summation to one. To ensure identifiability and prevent component ambiguity, we constrain the mean of the Gaussian component to exceed that of the exponential component in Equation~\eqref{equ-mean}. This configuration allows the exponential component to explicitly model the skewed density near zero, while the Gaussian components capture complex multimodal patterns in the mixture distribution. Following the formulation in Equation~\eqref{equ:egmd}, the complete mixture distribution conditioned on the input features is defined as:

\begin{equation}
\label{equ:condition-egm}
p(t|\mathbf{x}) = \omega_0(\mathbf{x})f_{\text{exp}}(t|\lambda(\mathbf{x})) + \sum_{k=1}^{K} \omega_k(\mathbf{x})f_{\text{gauss}}(t|\mu_k(\mathbf{x}), \sigma_k^2(\mathbf{x}))
\end{equation}

\subsection{Training Objective}

We optimize \name using a combination of three loss functions, each addressing different aspects of the prediction task.

\subsubsection{\textbf{Maximum Likelihood Estimation Loss}}

The primary objective is to maximize the likelihood \cite{pan2002maximum} of observed watch times under the \disname :

\begin{equation}
\mathcal{L}_{\text{MLE}} = -\frac{1}{N}\sum_{i=1}^{N}\log p(t_i|\mathbf{x}_i)
\end{equation}
where $N$ is the number of training instances, $t_i$ is the actual watch time for instance $i$, and $p(t_i|\mathbf{x}_i)$ is our model's predicted probability density at time $t_i$.

\noindent Expanding this using our mixture model definition in Equation~\eqref{equ:condition-egm}:

\begin{equation}
\begin{aligned}
\mathcal{L}_{\text{MLE}} = &-\frac{1}{N}\sum_{i=1}^{N}\log[\omega_0(\mathbf{x}_i)f_{\text{exp}}(t_i|\lambda(\mathbf{x}_i)) \\&+ \sum_{k=1}^{K} \omega_k(\mathbf{x}_i)f_{\text{gauss}}(t_i|\mu_k(\mathbf{x}_i), \sigma_k^2(\mathbf{x}_i))]
\end{aligned}
\end{equation}
This loss function guides the optimization process by encouraging our model to assign high probability density to the observed watch times. By minimizing this objective, \name learns to accurately capture the underlying distribution of user engagement patterns across different content types and viewing contexts.

\subsubsection{\textbf{Entropy Maximization Loss}}

To prevent the model from collapsing into a single component during training, we introduce an entropy maximization regularization term \cite{zhao2020domain} for the mixture weights:

\begin{equation}
\mathcal{L}_{\text{entropy}} = \frac{1}{N}\sum_{i=1}^{N}\sum_{k=0}^{K}\omega_k(\mathbf{x}_i)\log\omega_k(\mathbf{x}_i)
\end{equation}

Minimizing this loss is equivalent to maximizing the entropy between weights in $\omega(\mathbf{x}_i)$ and encourages the model to utilize multiple components when appropriate, rather than assigning all probability mass to a single component. This is crucial for maintaining the model's ability to capture the multimodal nature of the watch time distribution.

\subsubsection{\textbf{Regression Loss}}

To ensure that our model also performs well on absolute value estimation, we incorporate a regression loss between the expected value of \disname and the actual watch time:
\begin{equation}
\mathcal{L}_{\text{reg}} = \frac{1}{N}\sum_{i=1}^{N}|t_i - \hat{t}_i|
\end{equation}

\noindent where $\hat{t}_i$ is the expectation of $p(t|\mathbf{x})$ :

\begin{equation}
\label{eq:pred}
\hat{t}_i = \mathbb{E}[p(t|\mathbf{x}_i)] = \omega_0(\mathbf{x}_i)\frac{1}{\lambda(\mathbf{x}_i)} + \sum_{k=1}^{K}\omega_k(\mathbf{x}_i)\mu_k(\mathbf{x}_i)
\end{equation}

We use $\hat{t}_i $ as the final watch time prediction during the evaluation. By minimizing $\mathcal{L}_{reg}$, we explicitly optimize \name's ability to generate accurate prediction while maintaining its rich distributional awareness. Together with $\mathcal{L}_{MLE}$, this triple-objective design ensures that the model remains versatile in both probabilistic modeling and value regression.

\subsubsection{\textbf{Combined Loss Function}}

The final loss function is a weighted combination of the three individual losses:

\begin{equation}
\label{eq:total}
\mathcal{L} = \mathcal{L}_{\text{MLE}} + \alpha\mathcal{L}_{\text{entropy}} + \beta\mathcal{L}_{\text{reg}}
\end{equation}

\noindent where $\alpha$ and $\beta$ are hyperparameters controlling the relative importance of each objective. 
It is worth emphasizing that both the uniqueness of mixture model parameters in \name and the neural network's approximation capability to the optimal solution have sufficient theoretical guarantees \cite{mclachlan2019finite,celeux1995gaussian,panda2019deep}.

\subsection{Inference}

During inference, \name generates complete conditional probability distributions over possible watch time ranges in a fully end-to-end manner. This eliminates any need for computational preprocessing or reconstruction steps, making it particularly well-suited for industry deployments where computational efficiency and latency are critical factors.

For the standard watch time prediction, we simply utilize the expected value of $p(t|\mathbf{x})$ in Equation~\eqref{eq:pred} as an estimation. Moreover, \name has the flexibility to provide additional information: 

\begin{itemize}[leftmargin=*]
\item Recognition of quick-skipping behaviors.
\item Cumulative probability on specific watch time intervals.
\item Quantile estimation \cite{markovitch2002estimation} for specific user or video. 
\item Statistical confidence for uncertainty quantification \cite{soize2017uncertainty}.
\end{itemize}

In industrial recommender systems, this flexibility allows to leverage the same underlying model for diverse strategy designs without requiring specialized model variants.
\section{EXPERIMENTS}

In this section, we conduct extensive experiments and analyses to demonstrate the
effectiveness of the \name model. In these experiments, five research questions are explored:
\begin{itemize}[leftmargin=*]
\item RQ1: How does \name perform compared to existing state-of-the-art baselines in the offline/online watch time prediction task?
\item RQ2: To what extent can \name improve the recognition of quick-skipping behaviors compared to conventional approaches?
\item RQ3: How does the prior distribution selection contribute to overall effectiveness in predicting watch time?
\item RQ4: How does each part of the loss function(Equation~\eqref{eq:total}) affect the final performance of \name?
\item RQ5: How does our approach capture coarse-to-fine-grained distribution information in video watch time prediction?
\end{itemize}

\subsection{Experiment Settings}
\subsubsection{\textbf{Datasets}}
For offline experiments, we evaluate \name on a realistic industrial dataset (referred to as Indust) and three public datasets for recommender system. The details of these four datasets are as follows:
\begin{itemize}[leftmargin=*]
    \item Indust: An industrial dataset sampled from a real-world streaming short-video app with more than 200 million DAUs. We collect interaction logs for 14 days, which contain 951,870 interactions, 2,5947 users, and 7,155 videos. The sampling process ensures the consistency of the watch time distribution between the offline dataset and the online real-time distribution.
    \item KuaiRec\footnote{\url{https://kuairec.com/}}: A real-world dataset collected from the KuaiShou mobile app recommender logs \cite{gao2022kuairec}. It samples 12,530,806 impressions which cover 7,176 users and 10,728 videos.
    \item WeChat\footnote{\url{https://algo.weixin.qq.com/}}: This dataset is released by WeChat Big Data Challenge 2021, containing impression logs sampled from WeChat Channels within two weeks. It contains 7,310,108 interaction logs between 20,000 users and 96,418 videos.
    \item CIKM\footnote{\url{https://competitions.codalab.org/competitions/11161}}: A famous dataset sourced from the CIKM16 Cup Competition, is designed to predict user engagement duration in online search sessions. It contains 310,302 sessions and 122,991 items. Although CIKM is not derived from short video recommender systems, we still adopt it to prove the scalability of \name.
\end{itemize}
\subsubsection{\textbf{Offline Comparison(RQ1)}}
\label{sec:offline-comparison}
\begin{table}[]
    
    \setlength{\abovecaptionskip}{9pt}
    \setlength{\tabcolsep}{1.8pt}
    \small
    \caption{Offline comparison of watch time regression across different methods on four datasets. At least four significant digits are retained for each data point. The best improvements are highlighted in bold.}
    \label{tbl-offline}
    \centering
    \renewcommand{\arraystretch}{1.5}
    \begin{tabular}{c|cc|cc|cc|cc}
        
        \toprule
        \multirow{2}{*}{Method} & \multicolumn{2}{c|}{Indust} & \multicolumn{2}{c|}{KuaiRec} & \multicolumn{2}{c|}{WeChat} & \multicolumn{2}{c}{CIKM} \\
         & MAE  & XAUC & MAE & XAUC & MAE & XAUC & MAE & XAUC          \\ \hline
        \multicolumn{9}{l}{\textit{\textbf{Task: watch time regression}}} \\
        VR & 25.47&	0.6170&	4.730&0.5379	& 38.95	& 0.6009	&	1.3755&0.5946	 \\
        TPM & 25.22&	0.5998&	5.922&0.5510	& 21.47	& 0.6055	&	0.6435&0.7641 \\
        D2Q & 23.85&	0.6200&	4.583&0.5546	& 20.36	& 0.6270	&	0.6544&0.7597  \\
        CREAD & 25.58&	0.6245&	4.427&0.5821	& 19.21	& 0.6601	&	0.6363& 0.7713\\ 
        D$^2$CO &24.66&	0.6168&	5.087&0.5473	& 20.75	& 0.6257	&	0.6627&0.7595 \\ 
        \hline
        \name & \textbf{22.24} &	\textbf{0.6563} & \textbf{4.204}	 &	\textbf{0.6093} &	\textbf{18.88} & 	\textbf{0.6692} &	\textbf{0.6209} 	&\textbf{0.7751}  \\ 
        \bottomrule
\end{tabular}
\end{table}
\subsubsection{\textbf{Baselines}}
As comparison with \name, we choose several state-of-the-art methods for the watch time prediction task as baselines: Value Regression(VR), TPM  \cite{lin2023tree}, D2Q \cite{zhan2022deconfounding}, CREAD \cite{sun2024cread}, D$^2$CO \cite{zhao2023uncovering}. Comprehensive descriptions of the baseline models are provided in Section~\ref{sec:related} (Related Work).

\subsubsection{\textbf{Metrics}}
Following previous works \cite{zhan2022deconfounding,sun2024cread}, we adopt MAE and XAUC to measure the watch time prediction accuracy. Two metrics focus on absolute regression accuracy and ranking performance, respectively.
\begin{itemize}[leftmargin=*]
    \item MAE \cite{error2016mean} is measured as the average of the absolute error between the watch time prediction $\hat{y}_i$ and the ground truth $y_i$, which is formulated as:
\begin{equation}
    MAE = \frac{1}{N}\sum_{i=1}^{N}|\hat{y}_i - y_i|
\end{equation}
    \item XAUC \cite{zhan2022deconfounding} is used to measure the consistency between the predicted order and the ground truth order of the watch time value, which is formulated as:
\begin{equation}
    XAUC=\frac{1}{N(N-1)}\sum_{i=1}^{N}\sum_{j=1,j\neq i}^{N}\mathbb{1}((\hat{y}_i > \hat{y}_j) \oplus (y_i \leq y_j))
\end{equation}
    where $\mathbb{1}(\cdot)$ is the indicator function and $\oplus$ is the XOR operator. XAUC is estimated by randomly sampling pairs from the test set and checking whether their predicted order is consistent with the ground truth. A higher XAUC indicates better ranking performance. 
\end{itemize}

\subsubsection{\textbf{Implementation Details}}
\label{sec:impl-detail}
In the preprocessing of each dataset, we perform outliers filtering, sparse feature one-hotting, discretization/normalization of dense feature, etc. and $80\%$ of samples are used for training, while $20\%$ for evaluation. For three public datasets (KuaiRec, WeChat, CIKM), we follow the implementation of baselines\footnote{https://github.com/jackielinxiao/TPM} with random split on training set and evaluation set. In our industrial setting, we use temporal split for offline evaluation (Table1-Indust) and train EGMN on real-time temporal sample streams for online deployment (Table2). For fair comparison, we use the same setting for all methods. The MLPs in backbones use the same structure and the scale of trainable parameters for all methods is set at a similar magnitude. The embedding size of each feature is fixed to 16. The learning rate and mini-batch size are set to 0.1 and 2048, respectively. Training is done with Adagrad\cite{duchi2011adaptive} optimizer over shuffled samples. Throughout all experiments, we configure \name with 10 Gaussian components without meticulous parameter tuning. The loss weights $\alpha$ and $\beta$ in Equation~\eqref{eq:total} were set to 0.1 and 1.0, respectively. We open source all related code on Github\footnote{\url{https://github.com/BestActionNow/EGMN}}, including preprocessing of public datasets and implementation of all baselines.

\subsection{Main Results}
\label{sec:overall-performance}

In this section, We compare the offline watch time prediction effectiveness of \name with five baselines. The main experimental results are shown in Table~\ref{tbl-offline}. These results, showing an average improvement of 14.11\% in MAE and 7.76\% in XAUC across four datasets, demonstrate that the proposed \name model achieves state-of-the-art performance on the watch time prediction task on diverse platforms. Compared specifically to the second best performing model (predominantly CREAD), \name shows notable improvements of 6.75\% in MAE and 5.09\% in XAUC on Indust, 5.04\% and 4.67\% on KuaiRec, 1.72\% and 1.39\% on WeChat, and 2.42\% and 0.49\% on CIKM. The most substantial gains are observed on the Indust and WeChat datasets, where our model demonstrates consistent improvement across both metrics. These results conclusively validate the effectiveness of \name in capturing complex patterns for accurate watch time prediction across diverse datasets.

\begin{table}[]
\setlength{\abovecaptionskip}{9pt}
\caption{A/B testing results and online prediction accuracy of \name (proposed method) vs. CREAD (baseline), with bold values indicating superior metrics and red values denoting statistical significance  \cite{kohavi2013online,tang2010overlapping}.}

\label{tbl:online-2}
\setlength{\tabcolsep}{4pt}
\centering
\renewcommand{\arraystretch}{1.5}
\scalebox{0.9}{
\begin{tabular}{c|ccc|ccc}
\toprule
 \multirow{2}{*}{Method}&\multicolumn{3}{c|}{A/B Testing Results}& \multicolumn{3}{c}{Online Accuracy} \\ \cline{2-7}
  & \makecell[c] {Watch \\ Time} & \makecell[c] {Video \\ Views} & \makecell[c] {Engage \\ Actions}   & MAE & XAUC & KL \\ \hline
CREAD  & 11.75 & 37.00 & 1.824&  32.02 & 0.6826 &   0.1264   \\  
\name  & \textbf{11.83} & \textbf{37.07} & \textbf{1.823}& \textbf{31.37} &  \textbf{0.6912} & \textbf{0.1012} \\    
\hline
Diff.& \textcolor{red}{+0.681\%} & \textcolor{red}{+0.189\%} & -0.055\% & -2.030\% & +1.260\% & -19.94\% \\
\bottomrule

\end{tabular}}
\end{table}

As mentioned in Section~\ref{sec:impl-detail}, random split strategy is applied to three public datasets. It is worth noting that temporal split is indeed critical for industrial scenario to ensure online/offline consistency, thus we  also provide additional results for public datasets under the temporal split strategy: EGMN outperforms the second-best baseline with +1.77\% XAUC and -1.36\% MAE on KuaiRec, +2.41\% XAUC and -1.23\% MAE on WeChat, and +0.5\% XAUC with a significant -2.37\% MAE on CIKM. Along with the data in Table~\ref{tbl-offline}, it's verified that EGMN achieves SOTA performance regardless of the splitting strategy, demonstrating its strong generalization capability. 


\subsubsection{\textbf{Online A/B tests(RQ1)}}

To further demonstrate the effectiveness of \name, we conducted online A/B tests for 7 days on the industrial short video recommender system of Xiaohongshu App.  We simultaneously attribute 10\% real traffic to \name and our existing baseline CREAD. Our platform serves hundreds of millions of DAUs. Thus each traffic group covers enough real users and ensures the reliability of the experimental results.
\name and CREAD are applied to the ranking stage whose predicting backbone is MMOE \cite{ma2018modeling,tang2020progressive}. We show the results of online A/B testing in the left part of Table~\ref{tbl:online-2}. It is apparent that the proposed \name model achieves a  significant improvement (+0.681\%) in user watch time compared to the CREAD, while exhibiting no negative impact on other metrics. In addition, it is observed that \name also demonstrates a significant improvement (+0.189\%) in Video Views, which can be attributed to its better alignment with user interests and ultimate enhancement of user engagement.

It is well-known that online accuracy, compared to noise-free offline evaluations, provides a more reliable measure of model generalization in recommender systems. Consequently, alongside the A/B testing on user engagement metrics, we rigorously compare the real-time predictive accuracy between proposed \name and the baseline CREAD. As summarized in the right part of Table~\ref{tbl:online-2}, \name achieves significant improvements over CREAD in both MAE and XAUC, consistent with offline evaluations in Section~\ref{sec:offline-comparison}. Furthermore, we quantify the online alignment between predicted and actual watch time distributions via KL divergence \cite{kullback1951information}. \name reduces the KL divergence by nearly 20\% compared to CREAD, demonstrating  a closer approximation to the ground-truth distribution and directly validating its effectiveness in modeling real-time watch time distribution. 
\begin{figure}
\setcounter{figure}{2}
\setlength{\abovecaptionskip}{-2pt}
\setlength{\belowcaptionskip}{-8pt}
    \centering
    \scalebox{1.0}{
    \includegraphics[width=\linewidth]{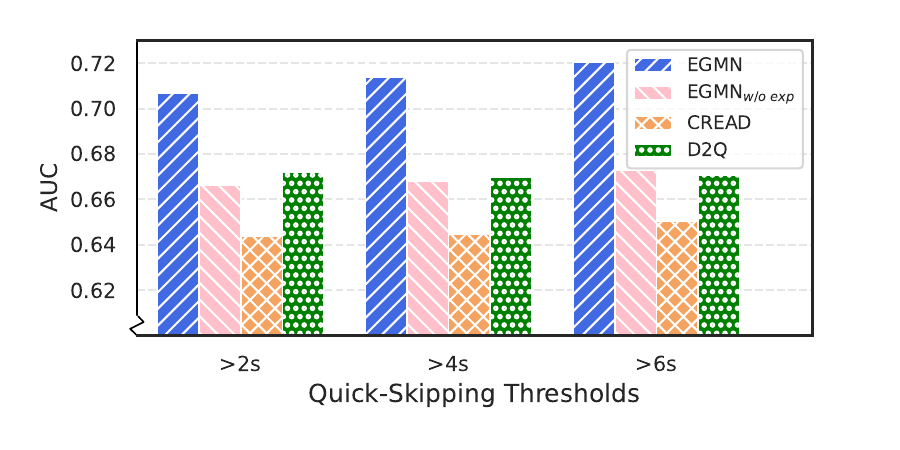}}
    \caption {AUC comparison among three binary classification tasks under different quick-skipping thresholds on Indust.}
    \label{fig:quick-skip_auc}
\end{figure}
\subsubsection{\textbf{Quick-Skipping Behavior Recognition(RQ2)}}
We evaluate the ability of \name to recognize quick-skipping behaviors which serve as an important implicit negative feedback signal in short-video recommender system. Figure~\ref{fig:quick-skip_auc} shows the discriminative power of several models in three quick-skipping thresholds (2s, 4s, and 6s). For each threshold, we formulate a binary classification task and compare the AUC scores \cite{huang2005using} of the predicted watch time. \name consistently outperforms CREAD and D2Q, achieving the highest AUC scores at all thresholds.

Moreover, we evaluate a variant of our model with the exponential component removed (\name\textsubscript{w/o exp}) in Figure~\ref{fig:quick-skip_auc}. A degradation exceeding 6\% in AUC is observed across all three quick-skipping thresholds, demonstrating that the exponential component within the \disname critically governs the capture of quick-skipping behaviors.

\subsection{Ablation Study}

Although Section~\ref{sec:overall-performance} shows a significant improvement in \name compared to other baselines, it is still necessary to investigate how each part of \name contributes to the total improvement. In this section, we conduct an ablation study on two datasets: Indust and KuaiRec.
Specifically, the ablation study is conducted from two perspectives: (1) Design of EGM Distribution  (2) Loss Function Selection. The experimental results are reported in Table~\ref{tbl:ablation}.

\subsubsection{\textbf{Design of EGM Distribution(RQ3)}} 
To evaluate the contribution of different components in \disname, we conduct ablation experiments by removing each component:

\textbf{Exponential Component}: Removing the exponential component leads to significant performance degradation, particularly on KuaiRec where MAE increases by 18.57\%. On Indust, we observe a more modest but still notable 3.06\% increase in MAE and a 2.29\% decrease in XAUC. Along with the results in Figure~\ref{fig:quick-skip_auc}, we attribute the degradation to the weakening of quick-skipping recognition. This further suggests that the exponential component plays a crucial role in modeling the coarse-grained distribution skewness.

\textbf{Gaussian Components}: Ablating the Gaussian components also degrades the model performance, with MAE increasing by 2.47\% on Indust and 9.71\% on KuaiRec. XAUC scores show corresponding decreases of 0.46\% and 5.05\%, respectively. To validate the contribution of Gaussian components for capturing the fine-grained distribution diversity, we further investigate the effect of Gaussian component number in \disname. Figure~\ref{fig:gasss} illustrates the impact of varying the number of Gaussian components in \disname. As shown, both the XAUC and MAE metrics exhibit a clear relationship with the Gaussian component number. The XAUC reaches its peak value of approximately 0.611 when using 8 Gaussian components, while MAE simultaneously achieves a relatively low value around 4.18. The results demonstrate that an insufficient number of Gaussian components limits the model's capacity to capture the complex patterns in watch time data. In contrast, an excessive number of components (beyond 12) leads to a significant decline in XAUC and an upward trend in MAE, suggesting overfitting issues and increased difficulty in optimizing the mixture model \cite{burnham2004multimodel}. In our offline and online practices, 8-12 Gaussian components approximately provide the best balance between regression accuracy and training stability.
\begin{table}[]
\setlength{\abovecaptionskip}{5pt}
\caption{Ablation study results of \name on two datasets: Indust and KuaiRec.}
\label{tbl:ablation}
\setlength{\tabcolsep}{1.8pt}
\centering
\small
\renewcommand{\arraystretch}{1.5}
\setlength{\abovecaptionskip}{-1.0cm}
\scalebox{0.9}{
\begin{tabular}{c|cc|cc|cc|cc}
\toprule
  \multirow{3}{*}{Ablated Method} & \multicolumn{4}{c|}{Indust} & \multicolumn{4}{c}{KuaiRec} \\ \cline{2-9}
   & \multicolumn{2}{c|}{MAE} & \multicolumn{2}{c|}{XAUC} & \multicolumn{2}{c|}{MAE} & \multicolumn{2}{c}{XAUC} \\ \cline{2-9}
   & value & diff & value & diff & value & diff & value & diff \\ \cline{1-9}
  
   \name & 22.24 & - & 0.6563 & -  & 4.204 & -  & 0.6093 & - \\ \hline
    \multicolumn{9}{l}{\textit{\textbf{Design of EGM Distribution:}}} \\

      $\ominus$ Exponential Comp. &   22.92 & +3.06\% & 0.6413 & -2.29\%  & 4.984  & +18.55\%  & 0.5945 & -2.43\% \\
    $\ominus$ Gaussian Comp. &   22.79 & +2.47\% & 0.6533 & -0.46\%  & 4.612  & +9.71\%  & 0.5785 & -5.05\% \\ \hline
    \multicolumn{9}{l}{\textit{\textbf{Loss Function Selection:}}} \\
     $\ominus\mathcal{L}_{MLE}$ &   23.07  &+3.73\%  & 0.6417 & -2.22\%  & 4.360 & +3.71\%  & 0.5928 & -2.71\% \\
    $\ominus\mathcal{L}_{entropy}$ &   22.95 & +3.19\% & 0.6555 & -0.12\%   & 4.461 & +6.11\% & 0.5862 & -3.79\%  \\
     $\ominus\mathcal{L}_{reg}$ &   22.96 & +3.24\% & 0.6534 & -0.44\%  & 4.253 & +1.17\%  & 0.6053 & -0.66\% \\

\bottomrule

\end{tabular}}
\vspace{-0.3cm}
\end{table}

\subsubsection{\textbf{Loss Function Selection(RQ4)}}
We also examine the impact of three loss functions in Equation~\ref{eq:total}:

\textbf{MLE Loss} ($L_{\text{MLE}}$): As the primary objective of mixed distribution fitting, $L_{MLE}$ plays a crucial role in unified modeling of duration distribution information across different granularity levels. Removing MLE loss in \disname significantly reduces the model's ability to capture complex distribution information, resulting in a negative effect on estimation accuracy, as shown in Table~\ref{tbl:ablation}.

\textbf{Entropy Loss} ($L_{\text{entropy}}$): Removing the entropy loss results in MAE increases of 3.19\% on Indust and 6.11\% on KuaiRec, with corresponding decreases in XAUC. A more in-depth analysis of this degradation has been performed, which confirms our hypothesis that $L_{entropy}$ effectively prevents the model from concentrating all weight on only one or two components, encouraging a more balanced utilization of the mixture distribution.

\textbf{Regression Loss} ($L_{\text{reg}}$): Without regression loss, performance deteriorates with MAE increases of 3.24\% on Indust and 1.17\% on KuaiRec. This shows that explicitly optimizing for expectation through regression remains important despite using a distribution-based approach.

Furthermore, we demonstrate the robustness of \name to hyperparameter variations. The sensitivity analysis reveals that both XAUC and MAE metrics remain relatively stable across different weight settings for entropy loss ($\alpha$) and regression loss ($\beta$): MAE remains stable between 22.13 – 22.90 (mean±std:22.34±0.252) and 22.08 – 23.05 (mean±std:22.62±0.305) for $\alpha$(0.2-2.0) and $\beta$(0.2-2.0) respectively. XAUC shows minimal variance which remains 0.6466 – 0.6583 (mean±std:0.6546±0.003) for $\alpha$(0.2-2.0) and 0.6496 – 0.6567 (mean±std:0.6538±0.002) for $\beta$(0.2-2.0).
The consistent performance across different hyperparameter settings indicates that \name is not overly sensitive to these loss weight configurations, suggesting good model stability and reducing the need for extensive hyperparameter tuning in practical applications.

In general, these ablation results validate our design choices for \name, showing that each component makes a meaningful contribution to the total performance. The mixed distribution approach with both exponential and Gaussian components, combined with our carefully designed loss functions, creates a robust framework for watch time prediction.

\begin{figure}
\setcounter{figure}{3}
\setlength{\abovecaptionskip}{10pt}
\setlength{\belowcaptionskip}{-14pt}
    \centering
    \scalebox{1}{
    \includegraphics[width=\linewidth]{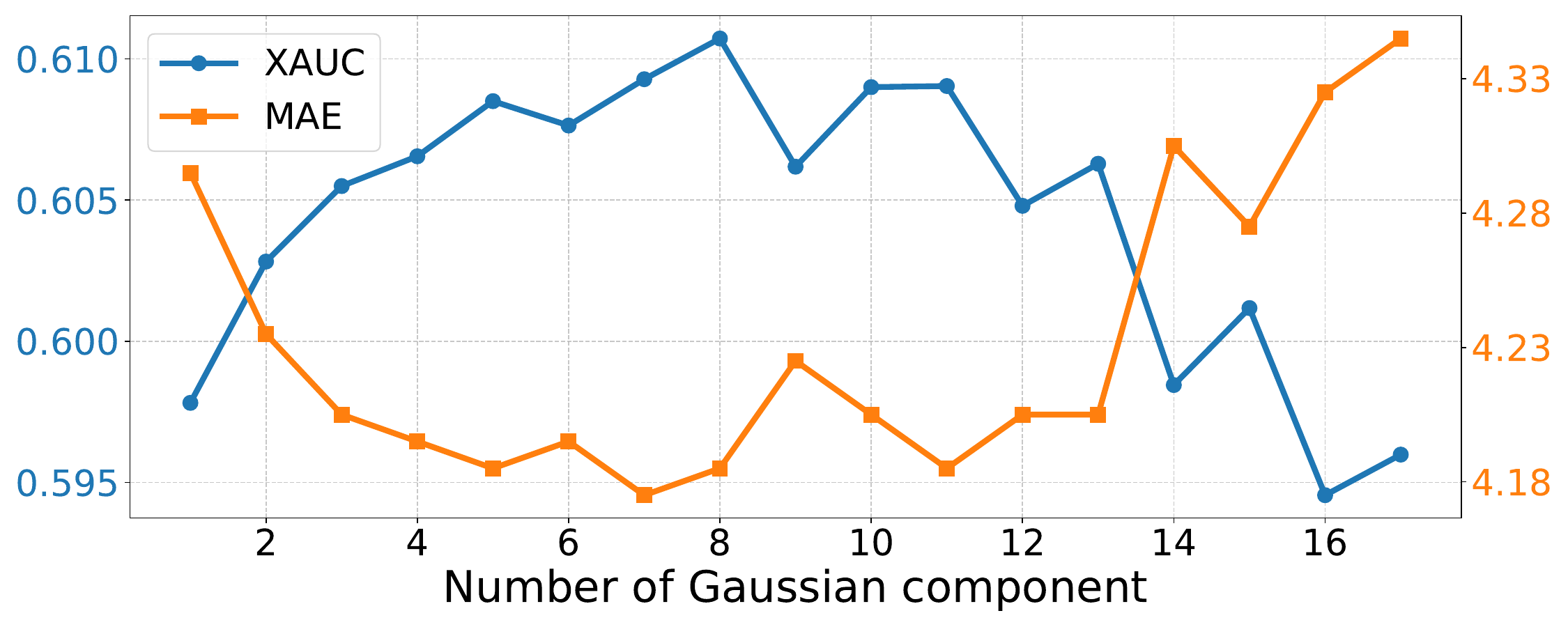}}
    \caption{Accuracy variation curve of \name with different numbers of Gaussian components on KuaiRec.}
    \label{fig:gasss}
    \vspace{-0.42cm}
\end{figure}

\subsection{Coarse-to-Fine-Grained Distribution Modeling(RQ5)}
\label{sec:coarse-to-fine}

Previous experiments have sufficiently demonstrated \name's superiority in value regression accuracy. In this section, we analyze \name’s distribution fitting capability to verify its effectiveness in capturing complex patterns within watch time distributions. Specifically, we evaluate \name against baselines at three granularity levels: overall distribution, distributions across various durations, and user-item-specific distributions, following a coarse-to-fine-grained paradigm as shown in Figure~\ref{fig:intro}.

\subsubsection{\textbf{Overall Distribution Modeling}}
\label{sec:dis-level1}

In this section, we compare the predicted watch time distributions of \name and other baselines against the actual distribution. Specifically, we calculate the KL divergence between model predictions and ground truth on the Indust dataset \cite{germain2015made,leipnik1991lognormal}. The results are shown in Figure~\ref{fig:kl}, where a lower KL divergence indicates better distribution fitting performance. It is worth noting that, except for \name and D2Q, all methods exhibit poor distribution fitting ability (KL divergence > 1.0), with VR failing to learn any meaningful patterns. \name converges to a KL divergence of 0.5, while D2Q achieves 0.1.

Comparative analysis of MAE trends between \name and D2Q is presented in the lower panel of Figure~\ref{fig:kl}. Notably, the MAE of D2Q exhibits an initial decline followed by a sharp increase after epoch 6, whereas \name maintains approximately monotonic descent. This contrast suggests that while D2Q's rigid constraint (directly fitting precomputed quantiles) achieves minimal KL divergence, it severely degrades value regression accuracy. Conversely, \name demonstrates joint convergence of KL divergence and MAE towards optima with synchronized dynamics throughout training epochs. In general, \name can better balance the trade-off between distribution modeling and value regression.

\subsubsection{\textbf{Duration-Level Distribution Modeling}}
\begin{figure}
\setcounter{figure}{4}
\setlength{\abovecaptionskip}{5pt}
    \centering
    \scalebox{1.0}{
    \includegraphics[width=\linewidth]{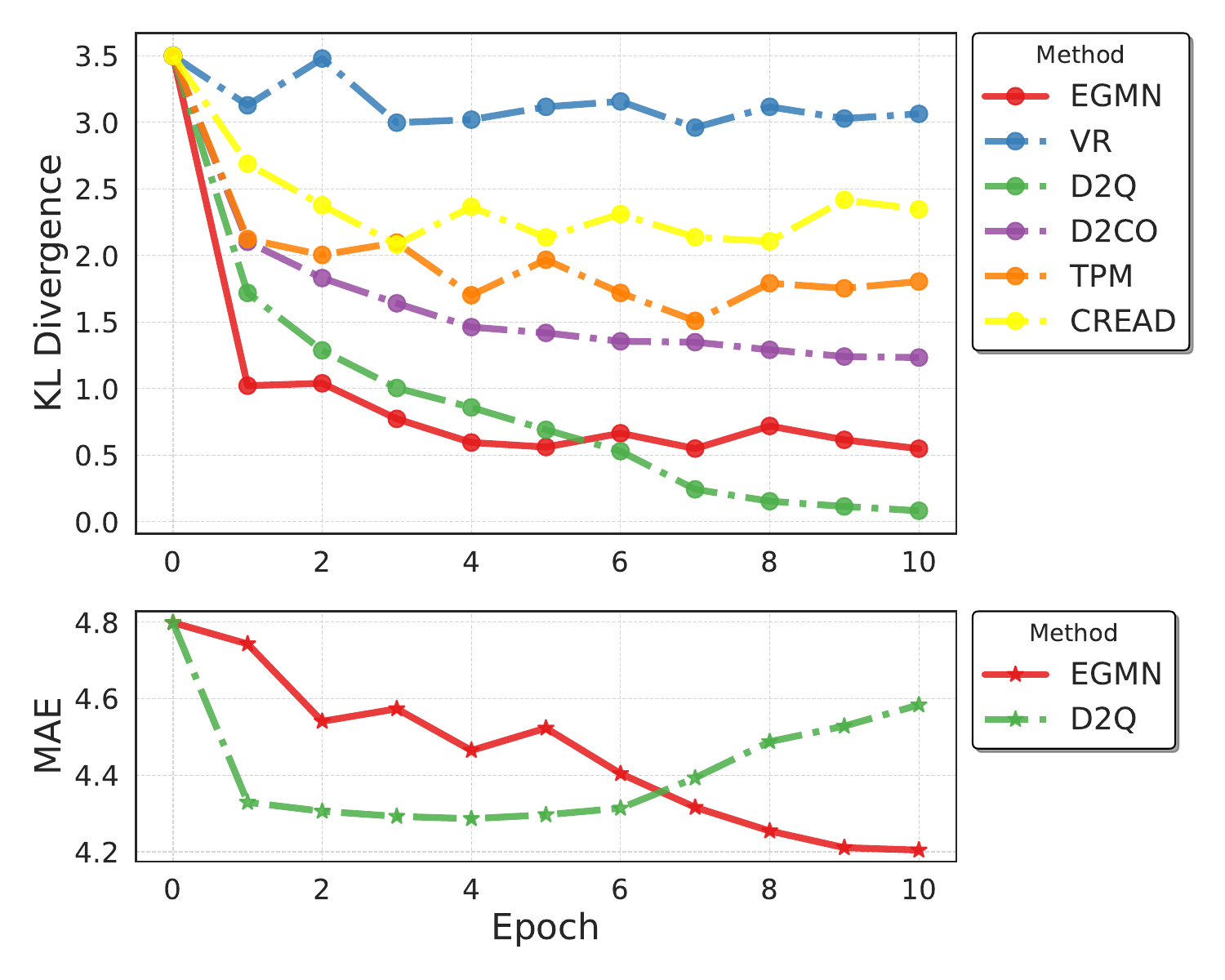}}
    \caption{Comparison of KL Divergence and MAE across different methods over training epochs on Indust.}
    \label{fig:kl}
    \vspace{-0.5cm}
\end{figure}

In this section, we compare the distribution modeling capability of \name and the baseline CREAD at the duration level. As shown in Figure~\ref{fig:dur}, we present the distributions of the actual watch time along with those estimated by \name and CREAD in four distinct duration intervals on the Indust dataset. As mentioned before, while the overall watch time distribution exhibits pronounced skewness around zero, the watch time distribution for videos sharing the same duration consistently displays bimodality: one peak near 3s (reflecting quick-skipping behavior) and another near the video duration (indicating full viewing completion). The estimated distribution of \name closely aligns with the ground truth, accurately preserving bimodal characteristics. Notably, \name achieves this without explicit duration debiasing operations, relying solely on duration as a model feature to capture duration-dependent distribution patterns. This also indicates that the hybrid architecture of \name inherently enables modeling of watch time distributions across other feature dimensions beyond duration. In contrast, CREAD’s estimation approximates Gaussian distributions across all duration intervals, with only minor mean shifts, indicating its inability to capture duration-level patterns.


\subsubsection{\textbf{User-Video-Level Distribution Modeling}}

\begin{figure}
\setcounter{figure}{5}
\setlength{\belowcaptionskip}{-10pt}
\setlength{\abovecaptionskip}{5pt}
    \centering
    \scalebox{1.0}{
    \includegraphics[width=\linewidth]{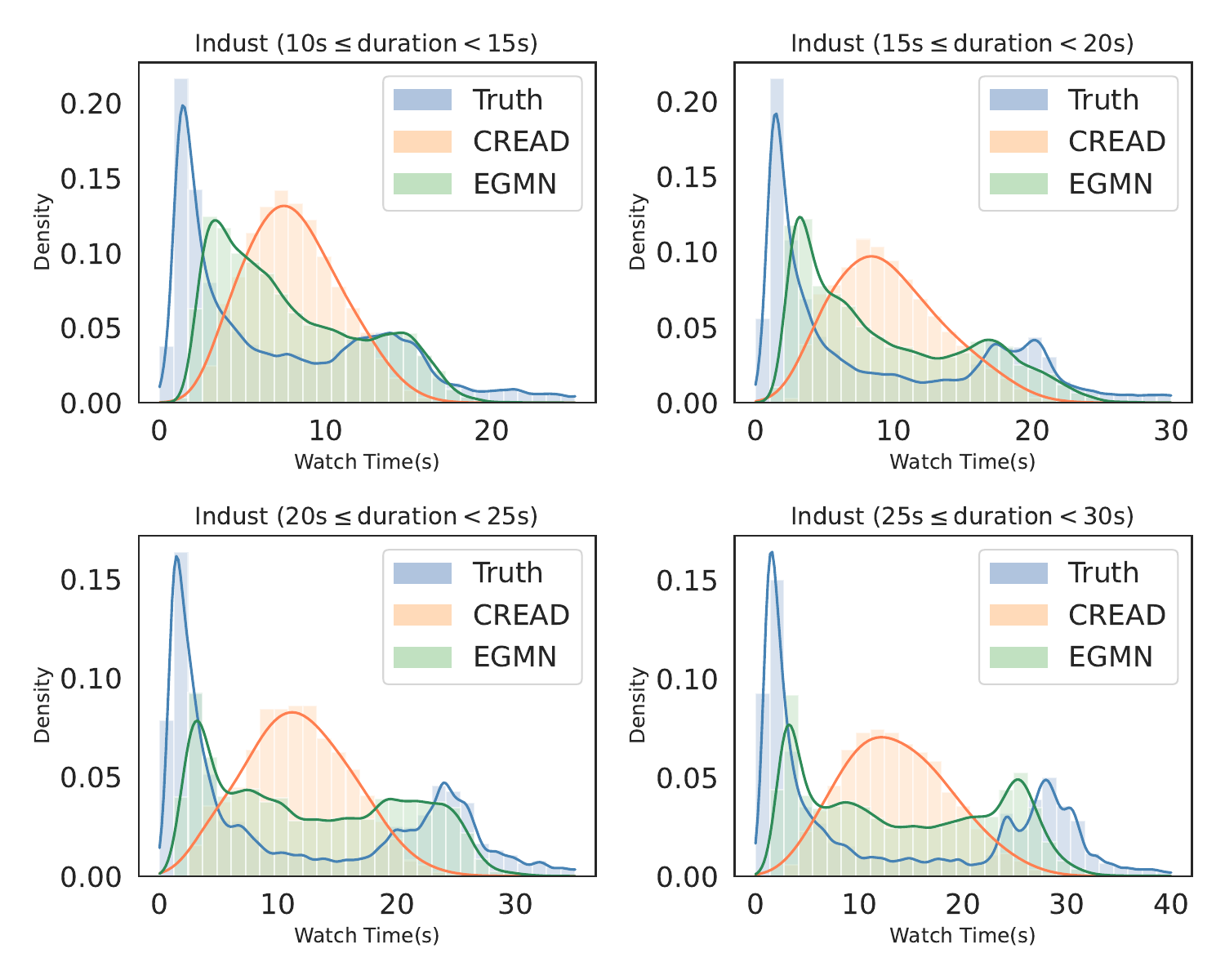}}
    \caption{Predicted vs. Actual watch time distribution across different duration intervals on Indust.}
    \label{fig:dur}
    \vspace{-0.5cm}
\end{figure}

To further evaluate \name's distribution modeling capability, we conduct a fine-grained assessment at the individual user-video interaction level, the most granular unit of analysis in recommender systems \cite{huang2023multi}. As mentioned in Section~\ref{sec:intro}, we select two representative users with distinct behaviors: User1 exhibits picking viewing patterns while User2 demonstrates more accepting viewing behavior. We also select two videos with different  types: Video1, a beauty-related video displaying a bimodal watch time distribution, and Video2, a movie compilation with a flatter, multi-peak distribution pattern.

In Figure~\ref{fig:user-item-dis}, the marginal distribution panels for User/Video(left column and top row) present histogram representations of the ground-truth probability density distributions for each user and video. The overlaid curves represent \name's predicted probability density. Notably, \name demonstrates remarkable fitting precision for both user and video marginal distributions, confirming its capability to model individual user consumption habits and video-specific engagement patterns accurately.

The $2\times2$ grid with yellow background illustrates how \name predicts watch time distributions for specific user-video pairs. These distributions reveal \name's ability to integrate information from both user behavior patterns and video characteristics. For example, examining $P(t|User2,Video2)$, we observe that the exponential component weight approaches zero, reflecting User2's less quick-skipping behavior, while the four peaks in this distribution correspond to the four peaks in Video2's marginal distribution.
This demonstrates \name's sophisticated capability not only to model separate user and video distributions but also to effectively combine these distributions to generate accurate joint prediction distributions. The model successfully captures the interaction effects between user preferences and video characteristics, resulting in predictions that reflect both the user's viewing habits and the video's inherent engagement patterns.

In summary, findings in Section~\ref{sec:coarse-to-fine} highlight \name's exceptional capacity for coarse-to-fine-grained distribution modeling, which is crucial for precise recommendations in systems where understanding the complete distribution of user watch time delivers significant value.


\begin{figure}
\setcounter{figure}{6}
\setlength{\abovecaptionskip}{10pt}
\setlength{\belowcaptionskip}{-2pt}
    \centering
    \scalebox{1.0}{
    \includegraphics[width=\linewidth]{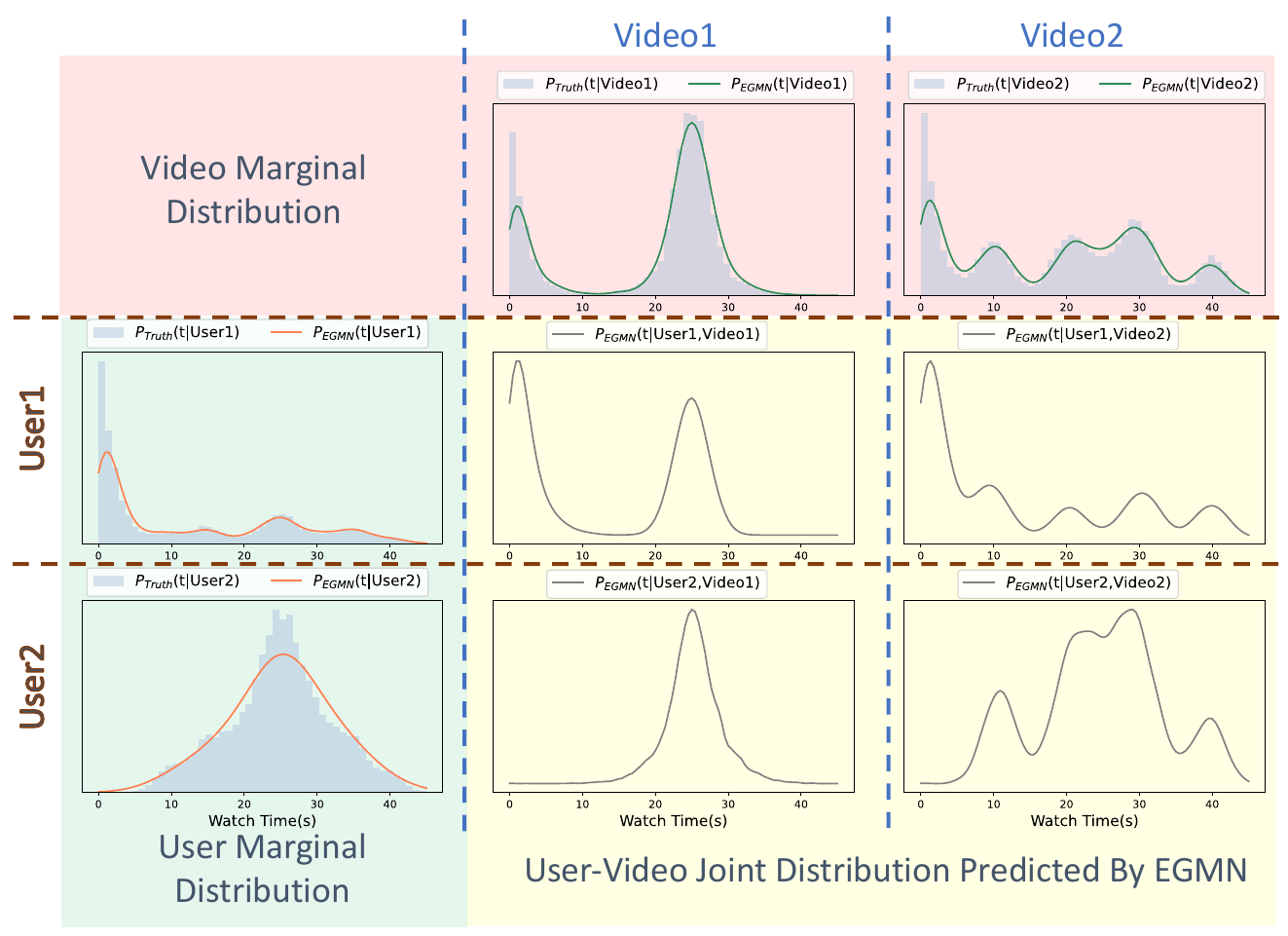}}
    \caption{Comprehensive evaluation of \name's distribution modeling capability at the user-video level.}
    \label{fig:user-item-dis}
    \vspace{-0.45cm}
\end{figure}

\vspace{10pt}
\section{CONCLUSION}
In this study, we address the critical challenge of accurate watch time prediction in short-video platforms by proposing a novel \name. By systematically analyzing real-world industrial data, we identify coarse-grained skewness and fine-grained diversity as key distributional obstacles. To tackle these challenges, we model watch time through an EGM distribution, where the exponential component captures the prevalent quick-skipping patterns, while the Gaussian component accounts for diverse user-video interaction behaviors. The proposed \name architecture effectively parameterizes this hybrid distribution. Extensive offline evaluations and online A/B tests on our large-scale industrial platform demonstrate that \name significantly outperforms state-of-the-art baselines, providing superior distribution fitting across multi-granularity levels. Notably, the success of \name highlights the importance of explicitly modeling complex real-world data distributions to improve prediction tasks. Considering that \name is proposed as a model-agnostic paradigm for the regression task, 
exploring the application of \name in other scenarios remains an exciting avenue for future work.

\bibliographystyle{ACM-Reference-Format}
\bibliography{custom}


\end{document}